\documentclass[11pt,a4paper]{article}
\pdfoutput=1

\usepackage{jcappub}
\usepackage[utf8]{inputenc}
\usepackage{soul}
\usepackage{graphicx}
\usepackage{verbatim}
\usepackage{epsfig}
\usepackage{subfigure}
\usepackage{color}
\usepackage{comment}
\usepackage{slashed}
\usepackage{amsmath}
\usepackage{ulem}

\newcommand{\be}{\begin{equation}} % only untightened
\newcommand{\ee}{\end{equation}}
\newcommand{\bea}{\begin{eqnarray}} % only untightened
\newcommand{\eea}{\end{eqnarray}}

\newcommand{\bmp}{\noindent\begin{minipage}{16cm}}
\newcommand{\emp}{\end{minipage}\vskip 7mm} % 7mm untightened
\def\lsim{\mathrel{\raise.3ex\hbox{$<$\kern-.75em\lower1ex\hbox{$\sim$}}}}
\def\gsim{\mathrel{\raise.3ex\hbox{$>$\kern-.75em\lower1ex\hbox{$\sim$}}}}
\newcommand{\ie}{{\it i.e.~}}

\newcommand{\intron}[1]{}%{#1}

\newcommand{\Sa}{\mathcal{S}_A}

\newcommand{\PkO}{\Delta^2_{\mathcal{R}}(k_0)}
\newcommand{\Pk}{\Delta^2_{\mathcal{R}}(k)}
\title{Axion dark matter from Higgs inflation with an intermediate $H_*$}

\author[a]{Tommi Tenkanen}
\author[b, c, d]{and Luca Visinelli}

\affiliation[a]{\,Department of Physics and Astronomy, Johns Hopkins University, \\
3400 N. Charles Street, Baltimore, MD 21218, USA}
\affiliation[b]{\,Department of Physics and Astronomy, Uppsala University, \\ L\"agerhyddsv\"agen 1, 75120 Uppsala, Sweden}
\affiliation[c]{\,Nordita, KTH Royal Institute of Technology and Stockholm University, \\ Roslagstullsbacken 23, 10691 Stockholm, Sweden}
\affiliation[d]{Gravitation Astroparticle Physics Amsterdam (GRAPPA), \\ Institute for Theoretical Physics Amsterdam and Delta Institute for Theoretical Physics,\\  University of Amsterdam, Science Park 904, 1098 XH Amsterdam, The Netherlands}

\emailAdd{ttenkan1@jhu.edu}
\emailAdd{luca.visinelli@physics.uu.se, l.visinelli@uva.nl}

\abstract{In order to accommodate the QCD axion as the dark matter (DM) in a model in which the Peccei-Quinn (PQ) symmetry is broken before the end of inflation, a relatively low scale of inflation has to be invoked in order to avoid bounds from DM isocurvature fluctuations, $H_* \lesssim \mathcal{O}(10^9)\,$GeV. We construct a simple model in which the Standard Model Higgs field is non-minimally coupled to Palatini gravity and acts as the inflaton, leading to a scale of inflation $H_* \sim 10^8\,$GeV. When the energy scale at which the PQ symmetry breaks is much larger than the scale of inflation, we find that in this scenario the required axion mass for which the axion constitutes all DM is $m_0 \lesssim 0.05{\rm \,\mu eV}$ for a quartic Higgs self-coupling $\lambda_\phi = 0.1$, which correspond to the PQ breaking scale $v_\sigma \gtrsim 10^{14}\,$GeV and tensor-to-scalar ratio $r \sim 10^{-12}$. Future experiments sensitive to the relevant QCD axion mass scale can therefore shed light on the physics of the Universe before the end of inflation.}

\begin{document}

\maketitle
%%%%%%%%%%%%%%%%%%%%%%%%%%%%%%%%%%%%%%%%%%%%%%%%%%%%%%%%%%%%%%%%%%%%%%%%%%%%%%%%%%%%%%%%%%%%%%%%%%%%

%%%%%%%%%%%%%%%%%%%%%%%%%%%%%%%%%%%%%%%%%%%%%%%%%%%%%%%%%%%%%%%%%%%%%%%%%%%%%%%%%%%%%%%%%%%%%%%

\section{Introduction}
\label{introduction}

The ``invisible'' QCD axion~\cite{Weinberg:1977ma, Wilczek:1977pj} is a light Goldstone boson arising upon the spontaneous breaking of a new global $U(1)$ symmetry, within the solution to the ``strong CP problem'' proposed by Peccei and Quinn (PQ)~\cite{Peccei:1977ur, Peccei:1977hh}. In addition, the QCD axion also serves as a well-motivated cold dark matter (CDM) candidate within a specific mass range~\cite{Abbott:1982af, Dine:1982ah, Preskill:1982cy}; for a review, see Ref.~\cite{Marsh:2015xka}. The level of precision reached in the assessment of the mass of the QCD axion and its dependence on the temperature of the QCD plasma from basic principles is quickly progressing, advancing in both extracting the QCD susceptibility from lattice computations~\cite{Borsanyi:2015cka, Bonati:2015vqz, Borsanyi:2016ksw, Petreczky:2016vrs} as well as in simulations of the string-wall network~\cite{Hiramatsu:2010yn, Hiramatsu:2010yu, Hiramatsu:2012gg, Klaer:2017qhr, Vaquero:2018tib, Gorghetto:2018myk, Buschmann:2019icd}.

The properties of the axion field and its evolution throughout the cosmological history of the Universe strongly depend on the relative energy scales associated with the breaking of the PQ symmetry, $v_\sigma$, and the Hubble rate at the end of inflation, $H_*$. In scenarios in which the scale of the PQ symmetry breaking is so high that the axion is a spectator field, \ie a field which remains energetically subdominant during inflation without taking part in causing the exponential expansion, and its energy density originates from its fluctuations during inflation, $H_* \ll v_\sigma$, the non-observation of primordial CDM density isocurvature by the \textit{Planck} collaboration~\cite{Ade:2018gkx, Akrami:2018odb, Aghanim:2018eyx} places strong constraints on model building, since the axion can be the dark matter particle only in models in which the energy scale of inflation is relatively low, $H_*  \lesssim \mathcal{O}(10^9)\,$GeV~\cite{Turner:1990uz, Beltran:2006sq, Hertzberg:2008wr, Visinelli:2009zm, Visinelli:2009kt, Visinelli:2014twa, Wantz:2009it, Visinelli:2017imh}. Recently suggested solutions include scenarios with a very low scale of inflation, even with $H_* \lesssim 1$ GeV \cite{Graham:2018jyp, Guth:2018hsa,Schmitz:2018nhb}.

Light QCD axions that spectate inflation have recently received attention both from the theory perspective~\cite{Turner:1990uz, Beltran:2006sq, Hertzberg:2008wr, Visinelli:2009zm, Visinelli:2009kt, Wantz:2009it, Visinelli:2017imh, Hoof:2018ieb} and from the perspective of detection in proposed and ongoing experiments~\cite{Irastorza:2018dyq}. In such a scenario, topological defects are washed out by inflation, and considerable additional effort to assess inhomogeneities like axion strings~\cite{Klaer:2017qhr, Klaer:2017ond, Ramberg:2019dgi} or axion miniclusters~\cite{Hogan:1988mp, Kolb:1993zz, Kolb:1993hw, Visinelli:2017ooc, Visinelli:2018wza} is not required. However, the challenge in this scenario relies on building a consistent theory of inflation that leads to a low enough scale of inflation, so that the axion can constitute all dark matter and simultaneously avoid the current bounds from the non-detection of dark matter isocurvature fluctuations. Other non-QCD axions that have been extensively considered in the literature as fields spectating inflation are the ``ultra-light'' axion particles which arise from string compactification, forming the so-called ``axiverse''~\cite{Svrcek:2006hf, Svrcek:2006yi, Arvanitaki:2009hb, Arvanitaki:2009fg, Higaki:2012ar, Cicoli:2012aq, Cicoli:2012sz, Higaki:2013lra, Stott:2017hvl, Visinelli:2018utg, Kinney:2018nny}, and which can possibly be detectable when invoking axion electrodynamics. The opposite regime in which inflation occurs at a scale much higher than the energy at which the spontaneous breaking of the PQ symmetry occurs, $H_* \gg v_\sigma$, has also received attention recently, with refined cosmological simulations set in the standard cosmological scenario yielding a narrow range in which the QCD axion would be the CDM particle~\cite{Klaer:2017ond, Gorghetto:2018myk, Buschmann:2019icd}.

In this work, we construct a simple mechanism to realise inflation with the Standard Model (SM) Higgs. In this scenario, the scale of inflation can be as low as $H_*\sim 10^8\,$GeV which is intermediate between what is usually obtained in single-field inflation models, $H_*\gtrsim 10^{13}\,$GeV, and what has been considered recently for axion models within a very low energy scale of inflation $H_* \sim 1$ GeV~\cite{Graham:2018jyp, Guth:2018hsa}. We thus consider the first of the two scenarios depicted above for which $H_* \ll v_\sigma$ and the QCD axion is a spectator field during inflation.

In this scenario, the Higgs field $\phi$ couples non-minimally to gravity via a term $\xi_\phi\,\phi^\dag \phi R$ in the Lagrangian, where $R$ is the Ricci scalar and $\xi_\phi$ is a dimensionless coupling constant. Such a setup is well-motivated, as non-minimal couplings between scalar fields and gravity are generated by quantum corrections in a curved background even if they are initially set to vanish at some scale~\cite{Birrell:1982ix}. As a result, the Higgs potential in the Einstein frame develops a plateau suitable for slow-roll inflation. The predictions of the model are in perfect agreement with the current observational data from the \textit{Planck} mission~\cite{Ade:2018gkx, Akrami:2018odb, Aghanim:2018eyx}, and the model has also been invoked in unified frameworks like the SMASH model~\cite{Ballesteros:2016xej}. However, the original Higgs inflation model~\cite{Bezrukov:2007ep} predicts a scale of inflation $H_* \gtrsim 10^{13}$ GeV which, due to the bound on CDM isocurvature perturbations, does not allow the axion to constitute all of the observed CDM~\cite{Turner:1990uz, Beltran:2006sq, Hertzberg:2008wr, Visinelli:2009zm}.

Instead, in this paper our approach is based on the so-called {\it Palatini} version of Higgs inflation~\cite{Bauer:2008zj} where, on top of the space-time metric, the connection is also assumed to be an {\it a priori} free variable. In the usual ``metric'' formulation of gravity, the space-time connection is determined by the metric only, \ie it is the usual Levi-Civita connection. Instead, in the Palatini formalism both the metric $g_{\mu\nu}$ and the connection $\Gamma$ are treated as independent variables, so that the Ricci scalar depends on both of them via $R\equiv g^{\mu\nu}R_{\mu\nu}(\Gamma)$, where the Ricci tensor is constructed from the Riemann tensor in the usual way. In the Palatini counterpart of Higgs inflation the scale of inflation can be as low as $H_*\sim 10^8\,$GeV (see Ref.~\cite{Almeida:2018oid}), allowing for the axion to successfully constitute all CDM without violating the CDM isocurvature bound. At the same time, the model predicts spectral features of temperature fluctuations in the Cosmic Microwave Background radiation (CMB) which are in perfect agreement with the data, as well as the preferred region for the axion to be the dark matter particle within the reach of upcoming detectors.

The paper is organised as follows. In Sec.~\ref{inflation}, we present the Lagrangian for the model and we revise how the Higgs can drive inflation when coupled non-minimally to gravity, whereas Sec.~\ref{axionDM} is dedicated for revising the isocurvature bounds on the axion field fluctuations. In Sec.~\ref{sec:observables} we present our main results for observables and provide further discussion. Finally, we draw our conclusions in Sec.~\ref{conclusions}.

%%%%%%%%%%%%%%%%%%%%%%%%%%%%%%%%%%%%%%%%%%%%%%%%%%%%%%%%%%%%%%%%%%%%%%%%%%%%%%%%%%%%%%%%%%%%%%%

\section{Description of the model}
\label{inflation}

\subsection{Decomposing the action}

In the so-called Jordan frame, the relevant part of the action describing the model is
\be
	S_J = \int d^4x \sqrt{-g}\left(-\frac{1}{2}\left(M_P^2 + 	F(\Phi) \right) g^{\mu\nu}R_{\mu\nu}(\Gamma) + \frac{1}{2} \delta_{IJ}g^{\mu\nu}\partial_{\mu}\Phi^I\partial_{\nu}\Phi^J - V(\Phi) \right) \,,
	\label{nonminimal_action1}
\ee
where $M_P$ is the reduced Planck mass, $g = {\rm det}(g^{\mu\nu})$, and $\Gamma$ is an {\it a priori} free connection appearing in the Ricci tensor $R_{\mu\nu}$. For simplicity, we assume that the connection is torsion-free, $\Gamma^\lambda_{\alpha\beta}=\Gamma^\lambda_{\beta\alpha}$. Throughout this paper, the particle physics sign convention $(+,-,-,-)$ is used, and we have indicated the space-time indices with Greek letters ($\mu$, $\nu$) and the field-space indices with capital letters ($I, J$), for both of which the Einstein summation convention is understood. We assume two complex scalar fields as the Higgs $\phi$ and the Peccei-Quinn $\sigma$ fields, so that $\Phi^1 = \phi$ and $\Phi^2 = \sigma$. We denote the scalar potential with $V(\Phi)$, while $F(\Phi)$ is a function that represents the non-minimal coupling between the $\phi$ and $\sigma$ fields and gravity and which will be specified below.

Within General Relativity (GR), the constraints imposed on the connection $\Gamma$ demand it to be the Levi-Civita connection, and hence renders the metric and the Palatini formalisms equivalent. However, when one considers non-minimally coupled matter fields or otherwise enlarged gravity sector, this is generally not the case~\cite{Sotiriou:2008rp}, and one has to make a {\it choice} of the underlying degrees of freedom in order to describe gravity. Currently, there are no reasons to favour one of the two formulations over the other one, and as we will see, allowing the connection to be independent of the metric does not amount to adding new degrees of freedom to the theory. However, the choice affects the field dynamics during inflation and hence also the predictions of the given model. This was originally noted in Ref.~\cite{Bauer:2008zj} and has recently gained increasing attention~\cite{Bauer:2010jg, Tamanini:2010uq, Rasanen:2017ivk, Tenkanen:2017jih, Racioppi:2017spw, Markkanen:2017tun, Jarv:2017azx, Racioppi:2018zoy, Enckell:2018kkc, Carrilho:2018ffi, Enckell:2018hmo, Antoniadis:2018ywb, Rasanen:2018fom, Kannike:2018zwn, Rasanen:2018ihz, Almeida:2018oid, Antoniadis:2018yfq, Takahashi:2018brt, Jinno:2018jei, Tenkanen:2019jiq, Rubio:2019ypq, Jinno:2019und, Giovannini:2019mgk} (see also Refs.~\cite{Azri:2017uor, Azri:2018gsz, Shimada:2018lnm, Aoki:2019rvi,Jimenez:2019ovq}). When the Higgs field relaxes to its electroweak vacuum after inflation, the usual Einstein-Hilbert gravity of GR is retained regardless of the choice of formalism, \ie metric or Palatini. Therefore, in our context gravity is modified only at early times by the presence of the non-minimal coupling between the Higgs field and gravity, just like in the original Higgs inflation model \cite{Bezrukov:2007ep}.

To ease the comparison with the original Higgs inflation~\cite{Bezrukov:2007ep} and the SMASH model~\cite{Ballesteros:2016xej}, we present the derivation of inflationary dynamics and the related observables in both metric and Palatini cases. In both cases, the non-minimal coupling in the Jordan frame action~\eqref{nonminimal_action1} can be removed by a Weyl transformation
\be \label{Omega1}
	g_{\mu\nu} \to \Omega^{2}(\Phi)g_{\mu\nu}, \hspace{1cm} \Omega^2(\Phi)\equiv 1+\frac{
	F(\Phi)}{M_P^2} \,,
\ee
which allows us to express the action for the Higgs field in the Einstein frame, in which the non-minimal coupling to gravity vanishes, as
\be \label{einsteinframe1}
	S_E = \int d^4x \sqrt{-g}\left(-\frac{1}{2}M_P^2\,R + \frac{1}{2} G_{IJ}(\Phi) g^{\mu\nu}\partial_{\mu}\Phi^I\partial_{\nu}\Phi^J - \frac{V(\Phi)}{\Omega^4(\Phi)} \right) \,,
\ee
where we have introduced the Ricci scalar $R = g^{\mu\nu}R_{\mu\nu}(\Gamma)$. In the Einstein frame, the scalars have acquired a non-trivial field-space metric given by
\be
	G_{IJ}(\Phi) = \frac{\delta_{IJ}}{\Omega^2(\Phi)} + \frac{3\kappa}{2}M_P^2\,\frac{\partial \ln\Omega^2(\Phi)}{\partial \Phi^I}\,\frac{\partial \ln\Omega^2(\Phi)}{\partial \Phi^J},
\ee
where $\kappa=1$ in the metric case and $\kappa=0$ in the Palatini case. Since the connection now appears only in the Einstein-Hilbert term, in the Einstein frame we have $\Gamma = \bar{\Gamma}$, \ie we retain the Levi-Civita connection 
\be
	\bar{\Gamma}^\lambda_{\alpha\beta} = \frac{1}{2}g^{\lambda\rho}(\partial_\alpha g_{\beta\rho} + \partial_\beta g_{\rho\alpha} - \partial_\rho g_{\alpha\beta}) \,,
\ee
which associates the metric uniquely with the connection. Therefore, with the conformal transformation we have transferred the dependence on the choice of gravitational degrees of freedom from the connection in the Jordan frame to the field-space metric in the Einstein frame.

\subsection{Higgs inflation and the QCD axion}

We expand the PQ field about the minimum $v_\sigma$,
\be
	\sigma(x) = \frac{1}{\sqrt{2}}\left(\rho(x) + v_\sigma\right)e^{-iA(x)/v_\sigma},
	\label{eq:PQfield}
\ee
where $\rho$ and $A$ are respectively the radial and angular modes, the latter being identified with the axion which is the Nambu-Goldsone boson of the theory. In the following, we parametrise the axion angle in terms of the axion field as $\theta(x) \equiv A(x)/v_\sigma$. However, as the axion field moves along a flat direction during inflation, in this section we neglect it, and instead focus on the radial mode $\rho$ only. The Higgs field values during inflation are much larger than the vacuum expectation value $v \approx 246\,$GeV, so that $\phi^{\dag} \simeq \left(0, h(x)\right)/\sqrt{2}$, in terms of the Higgs field $h\gg v$. The Weyl factor is then
\be
\label{eq:omega}
	\Omega^2(\Phi) = 1+\frac{F(\Phi)}{M_P^2} = 1 + \frac{\xi_\phi h^2 + \xi_\sigma(2\sigma^*\sigma - v_\sigma^2)}{M_P^2}\,,
\ee
where $\xi_\phi$ and $\xi_\sigma$ represent the dimensionless non-minimal coupling parameters of the $\phi$ and $\sigma$ fields with gravity. The change in the kinetic terms can be reabsorbed by redefining the Higgs field and the radial mode of the PQ field in terms of two fields $\chi$ and $\psi$, defined through~\cite{Almeida:2018oid, Takahashi:2018brt}
\bea
	\frac{d\chi}{dh} &=& \sqrt{\frac{1}{\Omega^2(h, \rho)} + \frac{6\kappa\, \xi_\phi^2}{\Omega^4(h, \rho)} \frac{h^2}{M_P^2}},\label{eq:chi}\\
	\frac{d\psi}{d\rho} &=& \sqrt{\frac{1}{\Omega^2(h, \rho)} + \frac{6\kappa\, \xi_\sigma^2}{\Omega^4(h, \rho)} \frac{v_\sigma^2}{M_P^2}} \,,\label{eq:psi}
\eea
which apply to first order in fluctuations. In this new parametrisation, the Einstein action in Eq.~\eqref{einsteinframe1} reads
\be \label{einsteinframe2}
	S_E = \int d^4x \sqrt{-g}\left[-\frac{M_P^2}{2}\,R + \frac{1}{2}g^{\mu\nu}\left(\partial_\mu \chi\partial_\nu \chi + \partial_\mu \psi\partial_\nu\psi + K_{\rm mix}\partial_\mu \chi\partial_\nu\psi\right) - \frac{V(\chi, \psi)}{\Omega^4(\chi, \psi)} \right] \,,
\ee
where the scalar potential is expressed explicitly in terms of the new fields $\chi(h)$ and $\psi(\rho)$, and we have introduced the kinetic mixing term
\be
	K_{\rm mix} = \frac{6\kappa\,\xi_\phi\xi_\sigma}{\Omega^4}\frac{hv_\sigma}{M_P^2}\frac{dh}{d\chi}\frac{d\rho}{d\psi} = \left(1+\frac{\xi_\phi}{6}\left(\frac{h}{\xi_\sigma v_\sigma}\right)^2\right)^{-1/2},
\ee
where the latter equality has been derived using Eqs.~\eqref{eq:omega}--\eqref{eq:psi}. We are interested in expressing the results in terms of Higgs inflation, for which we consider the flat direction of the potential along which the inflaton is mostly the Higgs, $\xi_\phi h^2\gg 6\xi_\sigma^2v_\sigma^2$ and $\rho\sim 0$. Therefore, we neglect the contribution from the kinetic mixing term. For a more exhaustive discussion on the topic, see Appendix A in Ref.~\cite{Almeida:2018oid}.

\begin{figure}[bt]
\begin{center}
	\includegraphics[width=.55\linewidth]{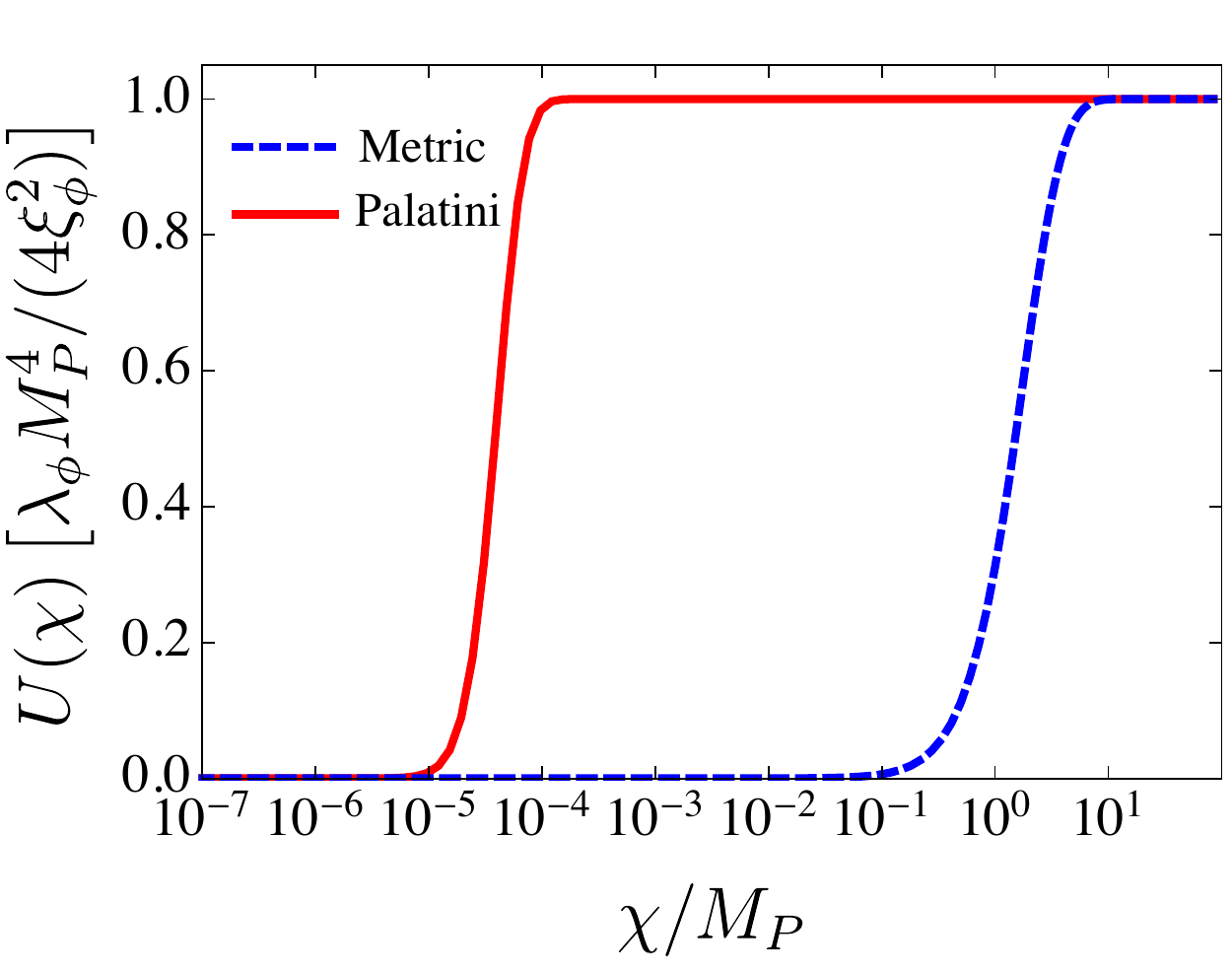}
	\caption{The Higgs potential $U(\chi)$ given in Eq.~\eqref{eq:quarticpotential} in units of $\lambda_\phi M_P^4/(4\xi_\phi^2)$, as a function of the field excursion $\chi/M_P$. Blue dashed curve: metric case. Red solid curve: Palatini case. Here we have fixed the non-minimal coupling $\xi_\phi = 10^9$ for both cases.}
	\label{potential_plot}
\end{center}
\end{figure}

The solution for the redefined Higgs field given in Eq.~\eqref{eq:chi} is then~\cite{GarciaBellido:2008ab,Bauer:2008zj,Rasanen:2017ivk}
\be
	\label{chi_solution}
	\sqrt{\xi_\phi}\frac{\chi}{M_P} = \sqrt{1+6\kappa\xi_\phi}\,{\rm arcsinh}\left(\sqrt{1+6\kappa\xi_\phi}\,u\right) - \kappa\sqrt{6\xi_\phi}\,{\rm arcsinh}\left(\frac{\sqrt{6\xi_\phi}u}{\sqrt{1+u^2}}\right) ,
\ee
where $u\equiv \sqrt{\xi_\phi}h/M_P$. Along this direction, we can treat the dynamics of the Higgs and the PQ field separately, once we have assumed that there is no coupling between the Higgs and the PQ fields in the scalar potential, in contrast to what has been considered in the SMASH model~\cite{Ballesteros:2016xej}. We integrate out the radial modes of the PQ field, so that the part of the action in Eq.~\eqref{einsteinframe1} that describes the Higgs field reads
\be
	S_{E, \chi} = \int d^4x \sqrt{-g}\left(-\frac{M_P^2}{2}R + \frac{1}{2}g^{\mu\nu}\,\partial_{\mu}\chi\,\partial_{\nu}\chi - U(\chi) \right) \,,
	\label{eq:action_phi}
\ee
where we have defined the Higgs potential
\be
	U(\chi) = \frac{\lambda_\phi}{4\Omega^4}h^4(\chi).
	\label{eq:quarticpotential}
\ee
Thus, under the conditions presented above, we retain the inflation model studied previously in Refs. \cite{Bauer:2008zj,Rasanen:2017ivk,Takahashi:2018brt}. We reiterate that when $\chi \to 0$, the usual Einstein-Hilbert gravity of GR is retained regardless of the choice of formalism (metric or Palatini), so gravity is modified only at early times by the presence of the non-minimal coupling. The canonically normalised field can be expressed as (see e.g. Ref. \cite{Takahashi:2018brt})
\bea
	h(\chi) \simeq
	\begin{cases} 
	\displaystyle\frac{M_P}{\sqrt{\xi_\phi}} \exp\left(\sqrt{\frac{1}{6}}\frac{\chi}{M_P} \right), & \quad \mathrm{metric} \,,\\   
	\displaystyle\frac{M_P}{\sqrt{\xi_\phi}}\sinh\left(\frac{\sqrt{\xi_\phi}\chi}{M_P}\right), & \quad \mathrm{Palatini} \,, 
	\end{cases}
\eea
and hence the large field Einstein frame potential reads
\bea
	U(\chi) = \frac{\lambda_\phi}{4\Omega^4}h^4(\chi) \simeq 
	\begin{cases}	
		\displaystyle\frac{\lambda_\phi M_P^4}{4\xi_\phi^2}
		\bigg[ 1 + \exp\left(-\sqrt{\frac{2}{3}} \displaystyle\frac{\chi}{M_P} \right) \bigg]^{-2},
		& \quad \mathrm{metric} \,,\\
		\displaystyle\frac{\lambda_\phi M_P^4}{4\xi_\phi^2}
		\tanh^4\left(\displaystyle\sqrt{\xi_\phi} \frac{\chi}{M_P}\right), & \quad \mathrm{Palatini} \,,
	\end{cases}
 	\label{chipotential1}
\eea
where the expressions in the metric case apply for $\xi_\phi \gg 1$ and $\chi \gg \sqrt{3/2}M_P$, whereas the expressions in the Palatini case are exact. In Fig.~\ref{potential_plot} we show the potential for the Higgs field as a function of the field excursion $\chi/M_P$, in the metric case (blue dashed curve) and in the Palatini case (red solid curve). We have expressed the potential in units of $\lambda_\phi M_P^4/(4\xi_\phi^2)$ and fixed $\xi_\phi = 10^9$, consistently with the findings obtained in previous studies~\cite{Bauer:2008zj,Rasanen:2017ivk,Takahashi:2018brt}. In the metric scenario, we have plotted the exact result for the quartic potential in Eq.~\eqref{eq:quarticpotential} with the field configuration from the solution~\eqref{chi_solution}, so that the plot for the blue curve is valid also in the region $\chi \ll \sqrt{3/2}M_P$. In both scenarios, the potential tends to a constant exponentially fast and is therefore suitable for slow-roll inflation. For the metric case, the field excursion is of the order of the Planck scale, while in the Palatini case we obtain $\chi \approx M_P/\sqrt{\xi_\phi} \ll M_P$. 

The inflationary dynamics is characterised by the slow-roll parameters
\bea
	\epsilon &\equiv& \frac{M_P^2}{2} \left(\frac{U'}{U}\right)^2\,, \label{SRparameters_epsilon} \\
	\eta &\equiv& M_P^2\, \frac{U''}{U} \,,\label{SRparameters_eta}
\eea
where a prime denotes differentiation with respect to $\chi$. For future convenience, we also introduce the total number of $e$-folds which parametrises the stretching of the scale factor during the inflationary period as
\be
	N_e  \equiv \frac{1}{M_P^2} \int_{\chi_f}^{\chi_i} {\rm d}\chi \, U \left(\frac{{\rm d}U}{{\rm d} \chi}\right)^{-1}\,.
	\label{Ndef}
\ee
Since inflation lasts as long as the slow-roll conditions $\epsilon \ll 1$ and $|\eta| \ll 1$ are satisfied, we define the value of the field $\chi_f$ at the end of inflation through the condition $\epsilon(\chi_f)= 1$. The field value $\chi_i$ is determined by a given $N_e$.

Before discussing observational consequences of the scenario, we make two remarks on the internal consistency of our results. First, all our considerations are based on a tree-level analysis. The computation of radiative corrections in non-minimally coupled theories is a subtle issue due to their intrinsic non-renormalizability and any sensible computation within the chosen framework requires the inclusion of an infinite number of higher-dimensional operators, which can be either associated to new physics or generated by the theory itself via radiative corrections~\cite{Bezrukov:2012hx} (see Ref.~\cite{Rubio:2018ogq} for a review), and which are therefore beyond the scope of this paper. Second, while concerns have been raised about the metric version of Higgs inflation suffering from violation of perturbative unitarity at scales relevant for inflation \cite{Burgess:2009ea,Barbon:2009ya,Barvinsky:2009ii,Bezrukov:2010jz,Burgess:2010zq,Hertzberg:2010dc}, in the Palatini case this is not a problem \cite{Bauer:2010jg}.

%%%%%%%%%%%%%%%%%%%%%%%%%%%%%%%%%%%%%%%%%%%%%%%%%%%%%%%%%%%%%%%%%%%%%%%%%%%%%%%%%%%%%%%%%%

\section{Axion dark matter}
\label{axionDM}

\subsection{Present axion abundance}

The distortion in the axion field due to the Weyl transformation has little influence on the dynamics of the radial component of the PQ field, which sits at the bottom of the PQ potential. On the other hand, the angular mode of the PQ field, the axion appearing in Eq.~\eqref{eq:PQfield}, evolves along the flat direction associated with the massless Nambu-Goldstone scalar field during inflation if the PQ symmetry is spontaneously broken during inflation. The part of the action in Eq.~\eqref{einsteinframe1} that describes the axion field only contains a kinetic term and thus describes a massless Nambu-Goldstone boson. However, an effective axion potential $V_{\rm QCD}(\theta, T)$ that depends on both the axion field configuration $\theta$ and temperature of the plasma $T$ arises due to the interaction of the PQ field with the QCD instantons around the QCD phase transition~\cite{Gross:53.43}. The action describing the QCD axion is
\be
	S_{E, \theta} = \int d^4x \sqrt{-g}\,\left[\frac{v_\sigma^2}{2}g^{\mu\nu}\partial_{\mu}\theta\partial_{\nu}\theta - V_{\rm QCD}(\theta, T)\right]\,,
	\label{eq:action_theta}
\ee
once we have neglected the interaction with the radial mode, which is a safe procedure when $v_\sigma \gg H_*$ and topological defects are not taken into account~\cite{Gorghetto:2018ocs}. The QCD potential grants a small mass to the axion which originates from non-perturbative effects during the QCD phase transition from mixing with the neutral pion~\cite{Weinberg:1977ma} as $m_0 = \Lambda^2/f_A$, with the theoretical expectation for the energy scale $\Lambda = 75.5\,$MeV and where the axion decay constant is $f_A = v_\sigma$. At higher temperature, the QCD instantons lead to an effective temperature-dependent potential $m_A(T)$, so that at zero temperature of the plasma we have $m_0 \equiv m_A(T=0)$. The values of the axion mass, self coupling and the effective potential $V_{\rm QCD}(\theta, T=0)$ have been computed at the next-to-leading order~\cite{Crewther:1979pi, DiVecchia:1981aev, diCortona:2015ldu, DiVecchia:2017xpu} starting from the QCD chiral Lagrangian~\cite{Weinberg:1978kz, Gasser:1983yg, Gasser:1984gg}. The temperature dependence of the axion potential has also been assessed through lattice computations~\cite{Bonati:2015vqz, Borsanyi:2015cka, Borsanyi:2016ksw, Petreczky:2016vrs, Dine:2017swf, Gorghetto:2018ocs}. 

We use these ingredients to compute the present abundance of non-relativistic axions which originated during an inflationary period. Computing the value of the present axion energy density is a standard procedure~\cite{Abbott:1982af, Dine:1982ah, Preskill:1982cy}, which usually assumes the conservation of the number of axions in a comoving volume from the onset of coherent field oscillations until present time. For this, we solve numerically for the axion field evolving in an expanding Friedmann-Robertson-Walker metric under the influence of the QCD axion potential $V_{\rm QCD}(\theta, T)$ and taking into account the change of the relativistic and entropy degrees of freedom with temperature. For the temperature dependence of the axion potential, we have used the results in Ref.~\cite{Borsanyi:2016ksw}. The exact numerical computation can be approximated within a factor of order two by the analytic estimate, leading to the present axion energy density
\be
	\rho_A(\theta_i) \approx \frac{3H_{\rm osc}}{2}\,m_0\,v_\sigma^2\left(\frac{a_{\rm osc}}{a_0}\right)^3\langle\theta_i^2\rangle\,,
	\label{eq:approx_density}
\ee
where the angle brackets denote spatial average. Here, the subscript ``osc'' refers to the moment when the QCD axion field starts to oscillate, so that the numerical result is well approximated by computing the temperature $T_{\rm osc}$ by setting $m_A(T_{\rm osc}) \approx 3H(T_{\rm osc})$. In order to compute the value of $H(T_{\rm osc})$, the temperature-dependence of the axion mass~\cite{Gross:53.43, Fox:2004kb}, the particle content of the underlying theory~\cite{DiLuzio:2016sbl, DiLuzio:2017pfr, DiLuzio:2017ogq}, and the underlying cosmology~\cite{Visinelli:2009kt, Visinelli:2017imh, Visinelli:2018wza, Ramberg:2019dgi} play a role. Assuming that around temperature $T_{\rm osc}$ the evolution of the Universe is described by the standard cosmological model and that the mass of the QCD axion scales as $m_A(T) \propto T^{-4}$, we obtain the present axion abundance as~\cite{Visinelli:2009kt}
\be
	\Omega_A h^2 = 0.754\,b\,\left(\frac{m_0}{\rm \mu eV}\right)^{-7/6}\langle\theta_i^2\rangle\,,
	\label{axion_abundance}
\ee
where $b$ is a factor of order one~\cite{Turner:1986} that captures the uncertainties coming from the approximation in computing $T_{\rm osc}$ instead of using the numerical solution. Our full numerical computation takes into account the anharmonic terms that are present in the axion potential and modify the prediction of the present abundance~\cite{Turner:1986, Linde:1987bx, Lyth:1992, Strobl:1994wk, Bae:2008ue, Visinelli:2009zm}. We can parametrise this effect  in the approximation in Eq.~\eqref{eq:approx_density} by writing the average over the initial axion angle $\theta_i$ as
\be
	\langle\theta_i^2\rangle \equiv F(\theta_i)\,\theta_i^2,
	\label{eq:anharmonic}
\ee
in terms of an anharmonic function $F(\theta_i)$. However, in presenting the numerical solution we do not make use of the function $F(\theta_i)$, which is shown here just for illustrative purposes.

\subsection{Axion isocurvature fluctuations}

Non-relativistic light bosons with masses $m\ll H_*$ that spectate inflation inherit primordial quantum fluctuations with an amplitude which is related to the energy scale of inflation $H_*$ and a nearly scale-invariant power spectrum described by
\be
	\frac{k^3}{2\pi^2}\langle |\delta A^2(k)|\rangle = \left(\frac{H_*}{2\pi}\right)^2\,.
\ee
For fields with $m\sim H_*$ the spectrum can have a non-trivial scale dependence; see Refs. \cite{Guth:2018hsa, Graham:2018jyp} in the context of axions and Refs. \cite{Starobinsky:1994bd,Markkanen:2019kpv} for a more general case. The primordial quantum fluctuations of the axion field later develop into isocurvature perturbations between axion dark matter and radiation~\cite{Axenides:1983, Linde:1985yf, Seckel:1985}, whose gauge-invariant definition is given by (see e.g. Ref. \cite{Wands:2000dp})
\be
	\Sa = \frac{3}{4}\frac{\delta\rho_\gamma}{\rho_\gamma} - \frac{\delta\rho_A}{\rho_A},
\ee
when both fluids satisfy their continuity equations $\dot{\rho_i}=3H(1+w_i)\rho_i$, where the overdot denotes derivative with respect to cosmic time, $w_i\equiv p_i/\rho_i$ is the effective equation of state parameter, and $p_i$ is the pressure of fluid $i = \gamma, A$. Here the perturbations of each fluid component are defined at point $x$ as $\delta \rho_i(x)\equiv \rho_i(x) - \langle \rho_i\rangle$. The axion density perturbations constitute an isocurvature mode because they are independent of the adiabatic perturbations seeded by the quantum fluctuations of the inflaton~\cite{Weinberg:2004kf}, and therefore totally uncorrelated with them. As the axions remain thermally decoupled from the rest of matter, the primordial isocurvature will not be washed away and can have observational consequences for the CMB and large scale structure formation.

In this framework, the standard deviation of the axion field in units of the decay constant is $\sigma_\theta = H_*/2\pi v_\sigma$. In the following, we assume that there are no couplings between the axion and the inflaton field other than gravity. Other scenarios have been discussed in Ref.~\cite{Ballesteros:2016xej} in relation to axion isocurvature fluctuations. Isocurvature fluctuations can also be suppressed by coupling the axion to a hidden sector~\cite{Kitajima:2014xla}, which we do not take into consideration here.

Assuming that axions constitute most of the observed CDM abundance $\rho_{\rm CDM}$~\footnote{The result in Eq.~\eqref{eq:axionisocurvatureboundth} gets modified in scenarios with mixed WIMP-axion dark matter~\cite{Bae:2014efa, Bae:2015rra, Baum:2016oow,Tenkanen:2019aij}.}, the power spectrum of axion isocurvature fluctuations is~\cite{Crotty:2003rz, Beltran:2005xd, Beltran:2006sq}
\bea
	\langle |\mathcal{S}_A|^2 \rangle &=& \left(\frac{\partial \ln \rho_A}{\partial \theta_i}\right)^2 \left(\frac{\rho_A}{\rho_{\rm CDM}}\right)^2\sigma_\theta^2 = \left(\frac{\rho_A}{\rho_{\rm CDM}}\right)^2\,\left(\frac{H_*}{\pi v_\sigma\,\theta_i}\right)^2\,\mathcal{F}(\theta_i)\,,
	\label{eq:axionisocurvatureboundth}\\
	\mathcal{F}(\theta_i) &=& \left(1 + \frac{\theta_i}{2}\frac{d}{d\theta_i}\ln F(\theta_i)\right)^2,
\eea
where the function $\mathcal{F}(\theta_i)$ is associated to the anharmonicity function $F(\theta_i)$ introduced in Eq.~\eqref{eq:anharmonic}. In our work, axion isocurvature fluctuations are evaluated using the first equality in Eq.~\eqref{eq:axionisocurvatureboundth}, using a numerical scheme to compute the derivative with respect to $\theta_i$ of the energy density. The latter equality is given for completeness to show the prediction of the analytic estimate, as we have pictured for the energy density by expressing the result in Eq.~\eqref{eq:approx_density}.

\section{Connection to observables}
\label{sec:observables}

Inflationary models generically predict the appearance of primordial scalar and tensor fluctuations, which redshift to super-horizon scales to later evolve into primordial perturbations in the density field as well as primordial gravitational waves, leaving an imprint in the CMB anisotropy and on the large-scale structure~\cite{Mukhanov:1981xt, Guth:1982ec, Hawking:1982cz, Starobinsky:1982ee, Bardeen:1983qw, Steinhardt:1984jj}. The spectrum of the adiabatic scalar perturbations generated during inflation is expressed by a power spectrum of primordial curvature perturbations $\Delta_{\mathcal{R}}^2(k)$ defined by~\cite{Kosowsky:1995aa, Leach:2002dw, Liddle:2003as}
\begin{eqnarray}
	\Pk \equiv \frac{k^3\,P_{\mathcal{R}}(k)}{2\pi^2} = \PkO\,\left(\frac{k}{k_0}\right)^{n_S-1}\,,
	\label{curvature_perturbations}
\end{eqnarray}
where $\PkO \sim 2.2\times 10^{-9}$ is the amplitude of the primordial scalar power spectrum~\cite{Ade:2018gkx, Akrami:2018odb} and $k_0$ is a typical scale at which the features of the spectrum are measured. We will take $k_0 = 0.002{\rm \,Mpc}^{-1}$ in the following. The scalar spectral index $n_S$ parametrises the mild dependence of the power spectrum on the co-moving wavenumber $k$. Single-field slow-roll inflation predicts that the scalar spectral index slightly deviates from the scale-invariant result $n_S = 1$, by a quantity that to the leading order depends on the slow-roll parameters as (see e.g. Ref.~\cite{Baumann:2009ds})
\begin{eqnarray}
	n_S - 1 \approx -6\epsilon + 2\eta\,.
	\label{eq:scalartilt}
\end{eqnarray}
In complete analogy, a power spectrum of tensor modes $\Delta_{\mathcal{T}}^2(k)$ is expected to be generated, which is observationally constrained by the tensor-to-scalar ratio at $k_0$,
\be
	r \equiv \frac{\Delta_{\mathcal{T}}^2(k_0)}{\Delta_{\mathcal{R}}^2(k_0)} = 16\epsilon.
	\label{eq:tsr}
\ee
The last expression in Eq.~\eqref{eq:tsr} is valid for single-field slow-roll inflation, so that along with Eq.~\eqref{eq:scalartilt}, the two observables $n_S$ and $r$ at the pivot scale $k_0$ are completely determined by the expressions for the slow-roll parameters Eqs. \eqref{SRparameters_epsilon} and \eqref{SRparameters_eta}. Measurements of the CMB temperature and polarisation anisotropies from the \textit{Planck} satellite augmented by the BICEP2/Keck Array (BK14) results set $n_S = 0.9653 \pm 0.0041$ (\textit{Planck} \textrm{TT, TE, EE + lowE + lensing + BK14} dataset combination at 68\% confidence level (CL)~\cite{Ade:2013zuv, Planck:2013jfk, Barkats:2013jfa, Ade:2015tva, Ade:2015lrj, Ade:2018gkx, Akrami:2018odb, Aghanim:2018eyx}), with small deviations of ${\cal O}(0.001)$ when independent data (such as Baryon Acoustic Oscillation distance measurements) are also included, or when different assumptions are made concerning the mass spectrum of massive neutrinos~\cite{Gerbino:2016sgw}. The same analysis of the dataset \textrm{TT, TE, EE + lowE + lensing + BK14} reports $r < 0.064$ at the $95\%$ CL~\cite{Ade:2018gkx, Akrami:2018odb, Aghanim:2018eyx}, measured at the quadrupole with $k_0 = 0.002{\rm \,Mpc}^{-1}$. In single-field slow-roll inflation, the Hubble expansion rate at the end of inflation $H_*$ is directly related to the measurements on the scalar power spectrum and the tensor-to-scalar ratio at the pivot scale $k_0$ as~\cite{Lyth:1984, Lyth:1992yy, Lyth:1998xn}
%
%\be
%	\frac{H_*}{M_P} = \sqrt{\frac{\pi^2\,\PkO\,r}{2}}\,.
%	\label{Hr}
%\ee
%
%
\be
	\PkO =  \frac{1}{2\epsilon}\left(\frac{\langle \delta\phi \rangle}{M_P}\right)^2 = \frac{2}{\pi^2\,r} \,\left(\frac{H_*}{M_P}\right)^2\,,
	\label{Hr}
\ee
where $\langle \delta\phi \rangle = H_*/2\pi$ describes the spectrum of fluctuations in the inflaton field and in the last expression we have used the relation $r = 16\epsilon$ which is valid for single-field slow-roll inflation. Since for Palatini inflation the field excursion is smaller by a factor $10^5$ with respect to the metric inflation, see Fig.~\ref{potential_plot}, in order to match the measured scalar power spectrum the first equality in Eq.~\eqref{Hr} requires a slow-roll parameter $\epsilon$ which is $10^{10}$ times smaller. For the scenario of a single-field inflation, this results in a tensor-to-scalar ratio which is also $10^{10}$ times smaller for the Palatini case compared to the metric case and therefore to a much lower Hubble scale of inflation.

The \textit{Planck} mission constraints axion isocurvature fluctuations by placing bounds on the primordial isocurvature fraction $\beta$, defined in terms of the power spectrum of isocurvature fluctuations at the scale $k_0$ as~\cite{Ade:2015xua, Akrami:2018odb}
\begin{equation}
	\Delta^2_A(k_0) = \langle |\mathcal{S}_A|^2 \rangle \equiv \Delta^2_{\mathcal{R}}(k_0)\frac{\beta}{1 - \beta}.
	\label{eq:axionisocurvaturebound}
\end{equation}
Using the combination of \textit{Planck} datasets \textrm{TT, TE, EE + lowE + lensing}, the fractional primordial contribution of uncorrelated dark matter isocurvature modes is constrained at the comoving wavenumber $k_0 = 0.002{\rm \,Mpc^{-1}}$ as $\beta < 0.035$ at 95\% CL~\cite{Ade:2015xua, Akrami:2018odb, Aghanim:2018eyx}.

Combining the expression for the scale of inflation in Eq.~\eqref{Hr} with the result in Eq.~\eqref{eq:axionisocurvatureboundth}, and using the bounds on the axion isocurvature fluctuations in Eq.~\eqref{eq:axionisocurvaturebound}, we obtain
\begin{equation}
	\frac{(v_\sigma\,\theta_i)^2}{\mathcal{F}(\theta_i)} = M_P^2\,\frac{1-\beta}{\beta}\,\frac{r}{2}\,.
	\label{eq:axionisocurvaturebound_p2}
\end{equation}
To express the importance of the result in Eq.~\eqref{eq:axionisocurvaturebound_p2} we stress that once we have taken into account that we are imposing $\rho_A = \rho_{\rm CDM}$, we obtain a relation between the initial misalignment angle $\theta_i$ and the scale at which the PQ symmetry spontaneously breaks, as $v_\sigma = v_\sigma(\theta_i)$. Using this result, the relation expressed in Eq.~\eqref{eq:axionisocurvaturebound_p2} reveals that the axion angle, and thus the value of the axion mass, are completely determined once both the tensor-to-scalar ratio $r$ and the primordial isocurvature fraction $\beta$ have been measured. A fit to the numerical solution for values $\theta_i < 0.1$ shows that the PQ symmetry breaking scale assumes a functional form $v_\sigma(\theta_i) = v_0\,\theta_i^\zeta$, where $v_0 = 1.5\times 10^{11}\,$GeV and the exponent $\zeta =-12/7$ has been recovered previously in the literature~\cite{Visinelli:2009zm}. This solution shows that in order to achieve a value $\theta_i = \mathcal{O}(0.1)$, the tensor-to-scalar ratio is expected to be $r \sim 10^{-12}$ for an isocurvature fraction $\beta = \mathcal{O}(10^{-2})$. In general, demanding that we wish to avoid a trans-Planckian value of the PQ symmetry breaking scale $v_\sigma \lesssim M_P$, leads to the stringent bound
\begin{equation}
	r \lesssim 10^{-8}\,\frac{\beta}{1-\beta}\,.
	\label{eq:newbound}
\end{equation}

We now discuss briefly how the bound obtained in Eq.~\eqref{eq:newbound} cannot be satisfied in the metric scenario, while it is easily accommodated within the Palatini approach. For this, we first derive the expressions for the tensor-to-scalar ratio and the scalar spectral index in these inflation models. Inserting the definitions of the slow-roll parameters in Eqs.~\eqref{SRparameters_epsilon} and~\eqref{SRparameters_eta} into Eqs.~\eqref{eq:scalartilt} and~\eqref{eq:tsr}, and expanding the expressions for the spectral index and the tensor-to-scalar ratio at $N_e \gg1$, gives 
\bea
	n_S(\chi_i) &\simeq& 1 - \frac{2}{N_e}\,,  \qquad\qquad \mbox{Metric and Palatini} \,,\label{eq:nS}\\
	r(\chi_i) &\simeq& 
	\begin{cases}
		\displaystyle\frac{12}{N_e^2},  & \quad \mathrm{Metric}\vspace{3mm} \,,\\
		\displaystyle\frac{24\pi^2\,\PkO}{\lambda_\phi N_e^4}, & \quad \mathrm{Palatini} \,. \\
	\end{cases}
	\label{eq:tensor_r}
\eea
Since the expression for the scalar spectral tilt in Eq.~\eqref{eq:nS} is the same for both the metric and Palatini approaches, the number of $e$-folds required to accommodate the measurements from Planck is of the order of $N_e \sim 50-60$ in both scenarios. To derive the above expressions, we have fixed $\xi_\phi$ according to the requirement of having the correct amplitude for the curvature power spectrum, so that the result in the Palatini case depends on $\lambda_\phi$ only (for details, see {\it e.g.} Ref.~\cite{Takahashi:2018brt}). Therefore, once a quartic coupling $\lambda_\phi = \mathcal{O}(0.1)$ is selected, Eq.~\eqref{eq:tensor_r} then predicts a value $r \sim 10^{-12}$ for the Palatini case, which is safely within the bounds in Eq.~\eqref{eq:newbound}, leading to a value of the axion misalignment angle $\theta_i = \mathcal{O}(1)$. On the contrary, the value of the tensor-to-scalar ratio predicted in the metric approach is too large to evade the bounds discussed above, since $r \sim 10^{-3}$ in this scenario regardless of $\lambda_\phi$ or $\xi_\phi$. This constitutes the main difference between the metric and Palatini scenarios and has been discussed exhaustively in Ref. \cite{Jarv:2017azx}. We also note in passing that due to the famous cosmological attractor behaviour \cite{Galante:2014ifa}, also the Starobinsky model in which inflation is driven by a scalar degree of freedom contained in an $R^2$ term in the action predicts a similar value for $r$ as the metric case \cite{Starobinsky:1980te}, and is therefore ruled out in the present context where the QCD axion constitutes the CDM.

\begin{figure}[bt]
\begin{center}
	\includegraphics[width=.7\linewidth]{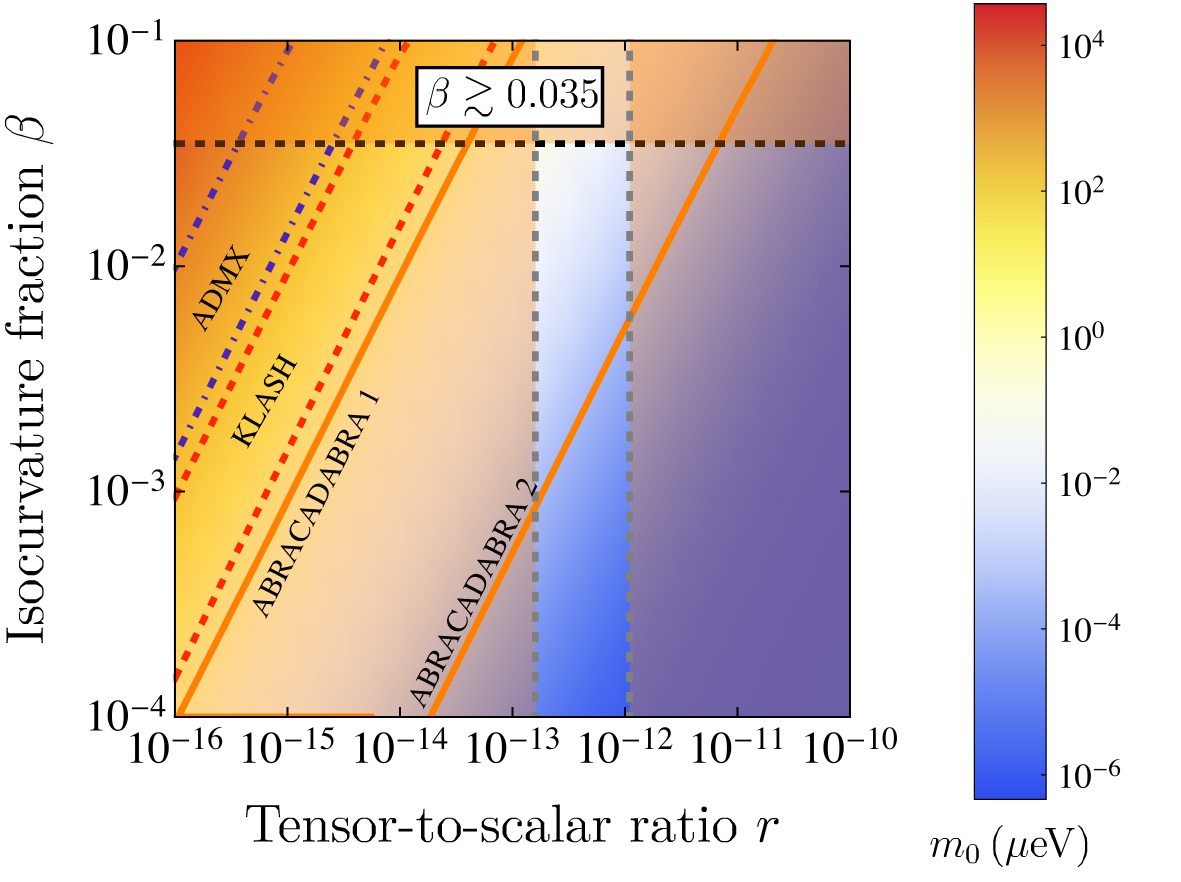}
	\caption{The value of the axion mass (in ${\rm \mu eV}$, colour scale to the right of the figure) that yields the observed dark matter abundance as a function of the tensor-to-scalar ratio $r$ (horizontal axis) and the primordial isocurvature fraction $\beta$ (vertical axis), as obtained from Eq.~\eqref{eq:axionisocurvaturebound_p2}. The horizontal black dashed line marks the region excluded by the non-observation of $\beta \gtrsim 0.035$ at 95\% CL from \textit{Planck}, while the vertical dashed lines mark the regions excluded by the measurement of the scalar spectral tilt $n_S$ when assuming Higgs inflation within the Palatini approach and a quartic Higgs self-coupling $\lambda_\phi = 0.1$. We also show the sensitivity that is expected to be reached by ABRACADABRA in configurations 1 and 2 (orange thick lines, see text), KLASH (within the red dashed lines), and ADMX (within the blue dot-dashed lines).}
	\label{fig:densityplot}
\end{center}
\end{figure}

Therefore, we now focus on the Palatini scenario only, which as we have shown above is the natural stage to set the study of axion CDM produced during inflation, for which a relatively low energy scale is required in order to evade the bounds from the non-observation of primordial dark matter isocurvature~\cite{Turner:1990uz, Beltran:2006sq, Hertzberg:2008wr, Visinelli:2009zm, Visinelli:2009kt, Wantz:2009it, Visinelli:2017imh}. Combining Eq.~\eqref{Hr} with the expression for the tensor-to-scalar ratio for the Palatini scenario in Eq.~\eqref{eq:tensor_r}, we obtain a value of the Hubble rate at the end of inflation
\begin{equation}
H_*\simeq 2\times 10^8\,{\rm GeV}\,, 
\end{equation}
for the choices  $\lambda_\phi \sim 0.1$ and $N_e = 52$, which is in accord with Refs. \cite{Takahashi:2018brt,Rubio:2019ypq} where $N_e\simeq 50$ for $k_0 = 0.05{\rm \,Mpc^{-1}}$ was found. Similarly, inserting the expression in Eq.~\eqref{eq:tensor_r} into Eq.~\eqref{eq:axionisocurvaturebound_p2} gives a relation between the axion misalignment angle $\theta_i$ and the primordial isocurvature fraction $\beta$ which is valid in the Palatini scenario, as
\be
	\frac{\beta}{1 - \beta} = \frac{12\pi^2\,\PkO}{\lambda_\phi\,N_e^4} \left(\frac{M_P}{v_\sigma(\theta_i)\,\theta_i}\right)^2\,\mathcal{F}(\theta_i)\,,
\ee
where $v_\sigma = v_\sigma(\theta_i)$ because of the assumption that the axion is the dark matter particle. A second relation, valid in the Palatini scenario, is obtained by eliminating the number of $e$-folds $N_e$ appearing in Eqs.~\eqref{eq:nS} and~\eqref{eq:tensor_r}, which we reduce to a single expression as
\be
	n_S = 1 -  \left(\frac{2\lambda_\phi\,r}{3\pi^2\PkO}\right)^{1/4}\,.
	\label{eq:ns_r}
\ee
The product $\lambda_\phi\,r$ is then constrained by demanding that the scalar spectral tilt lies within the 95\% CL region as measured by the \textit{Planck} mission, using the result in Eq.~\eqref{eq:ns_r}, as
\be
	1.6 \times 10^{-14} \lesssim \lambda_\phi\, r \lesssim 1.1 \times 10^{-13}\,.
	\label{eq:bounds}
\ee
Results in Fig.~\ref{fig:densityplot} show the value of the axion mass (the colour scale to the right of the figure in $\mu$eV) in the plane $(r,\beta)$, according to the expression in Eq.~\eqref{eq:axionisocurvaturebound_p2}. The value of the axion mass on the colour scale is bounded from above by astrophysical considerations on the cooling time of stellar objects~\cite{Vysotsky:1978dc, RAFFELT1986402, Berezhiani:1992rk, Raffelt2008, Viaux:2013lha, Giannotti:2017hny}, while the lower bound corresponds to demanding $v_\sigma < M_P$. Note that the colour chart for the axion mass does not depend on the inflation model considered, since the derivation in Eq.~\eqref{eq:axionisocurvaturebound_p2} is a general result within single-field, slow-roll inflation. However, the result assumes that the axion is the cold dark matter particle, and that the PQ symmetry is broken during the inflationary epoch, so that the nearly massless axion acquires the isocurvature fluctuations described by the power spectrum in Eq.~\eqref{eq:axionisocurvaturebound}. The horizontal bound limits the parameter space to the region $\beta \lesssim 0.035$, which is the 95\% exclusion region inferred by the \textit{Planck} mission~\cite{Ade:2015xua, Akrami:2018odb, Aghanim:2018eyx}. The vertical lines show the bounds obtained from using Eq.~\eqref{eq:bounds} with $\lambda_\phi = 0.1$, and give the model-dependent constraints to the parameter space. 

We emphasise that the results~\eqref{eq:ns_r} and~\eqref{eq:bounds} are obtained within the Palatini scenario. As discussed above, deriving the same bounds in the metric inflation scenario would lead to a different window in which $r$ is allowed, centred around the value $r \sim 10^{-3}$ and thus disfavoured by the trans-Planckian value of the PQ energy scale inferred. The vertical bounds are thus dependent on the model considered, and are here derived for the Palatini scenario with a quartic Higgs potential and $\lambda_\phi = 0.1$. In this scenario, the predicted value of the axion mass is $m_0 \lesssim 0.05{\rm \, \mu eV}$, corresponding to $v_\sigma \gtrsim 10^{14}\,$GeV, which can be partially probed by the ABRACADABRA experiment~\cite{Kahn:2016aff, Ouellet:2018beu} when operating in the ``broadband'' configuration within a cavity of magnetic field $B_0 = 5\,$T and a volume $V = 1{\rm \,m}^3$ (``ABRACADABRA 1'' in the figure), covering the mass range down to the line marking when operating in either the ``resonant'' or the ``broadband'' configuration, within a cavity of magnetic field $B_0 = 5\,$T and a volume $V = 100{\rm \,m}^3$ (``ABRACADABRA 2'' in the figure). In Fig.~\ref{fig:densityplot}, we also show the expected reach of the haloscope searches~\cite{Sikivie:1983ip, Sikivie:1985yu, Graham:2015ouw, Irastorza:2018dyq} by ADMX~\cite{Duffy:2006aa, Asztalos:2009yp, Asztalos:2011bm, Stern:2016bbw} and KLASH~\cite{Alesini:2017ifp}, which are also going to explore values of the axion mass that are lighter than what has been recently inferred by numerical computations of the dynamics of the PQ field~\cite{Klaer:2017qhr, Klaer:2017ond, Gorghetto:2018myk, Vaquero:2018tib, Buschmann:2019icd} in a different cosmological scenario. On a theoretical viewpoint, the results obtained do not depend on the coupling of the QCD axion with the photons~\cite{Kim:1979if, Shifman:1979if, Zhitnitsky:1980tq, Dine:1981rt}, which is nonetheless present when considering the experimental setup~\cite{Wilczek:1987mv, Itin:2007wz, Visinelli:2013fia, OHare:2017yze, Knirck:2018knd, Visinelli:2018zif}. The limits drawn in Fig.~\ref{fig:densityplot} are for the KSVZ axion scenario~\cite{Kim:1979if, Shifman:1979if}.

In the future, the EUCLID satellite will constrain $\beta$ roughly at a percent level \cite{Amendola:2016saw}, which places constraints to different dark matter models regardless of the underlying model of inflation. In contrast to the metric approach, for which the predicted value $r \sim 10^{-3}$ is well within reach of current and planned future experimental searches such as LiteBIRD~\cite{Matsumura:2013aja}, CORE~\cite{Remazeilles:2017szm}, and the Simons Observatory~\cite{Simons_Observatory}, the Palatini approach leads to a sensibly small value of the tensor-to-scalar ratio $r \sim 10^{-12}$ which is not possible to be probed directly with any near future experiment. For this reason, the discovery of a light axion of mass $m \lesssim 0.1\,\mu$eV could provide evidence for the Palatini-Higgs inflation scenario, in which the scale of inflation and the associated value of the tensor-to-scalar ratio are sufficiently low as to satisfy the inequality in Eq.~\eqref{eq:newbound}. In fact, in the Palatini scenario we have predicted $H_* \sim 10^8$ GeV for $\lambda_\phi \sim 0.1$ and $N_e \sim 50$, which is exactly what is required for the axion field to constitute all DM in the above mass range.

%%%%%%%%%%%%%%%%%%%%%%%%%%%%%%%%%%%%%%%%%%%%%%%%%%%%%%%%%%%%%%%%%%%%%%%%%%%%%%%%%%%%%%%%%%

\section{Conclusions}
\label{conclusions}

In this paper we have constructed a particularly simple model in which the Standard Model Higgs field is non-minimally coupled to gravity and acts as the inflaton, leading to a scale of inflation $H_* \sim 10^8\,$GeV when Palatini gravity is assumed. When the PQ symmetry is incorporated in the model and the energy scale at which the symmetry breaks is much larger than the scale of inflation, we found that in this scenario the required axion mass for which the axion constitutes all DM is $m_0 \lesssim 0.05{\rm \,\mu eV}$ for a quartic Higgs self-coupling $\lambda_\phi = 0.1$, which correspond to the PQ breaking scale $v_\sigma \gtrsim 10^{14}\,$GeV and tensor-to-scalar ratio $r \sim 10^{-12}$. The model avoids all isocurvature constraints and can be tested in large parts with future experiments sensitive to the QCD axion mass above.

We reiterate that as within the General Relativity the metric and Palatini formalisms are equivalent, currently there are no reasons to favour one theory of gravity over another in the context where the inflaton field couples non-minimally to gravity but decays away after inflation, thus retaining GR at late times. However, what makes the Palatini scenario particularly interesting is the possibility for $r$ taking a small value, even $\mathcal{O}(10^{-12})$, whereas in the metric case it is always bound to values $\mathcal{O}(10^{-3})$. As we showed, this aspect is crucial for the QCD axion to constitute all dark matter.

The model presented in this paper is not only very successful but also simple. Other models in which the tensor-to-scalar ratio can be drastically suppressed consist of Natural inflation~\cite{Freese:1990rb, Adams:1992bn, Savage:2006tr} in the regime of warm inflation scenario~\cite{Berera:1995ie}, {\it i.e.} the so-called Natural Warm Inflation scenario~\cite{Visinelli:2011jy, Visinelli:2014qla, Visinelli:2016rhn}, or models where the gravity sector includes also a Starobinsky-type $R^2$ term in the Palatini formulation \cite{Enckell:2018hmo,Antoniadis:2018ywb, Antoniadis:2018yfq,Tenkanen:2019jiq}. It may also be possible to suppress the tensor-to-scalar ratio by incorporating quantum corrections to a given inflationary model. It would be interesting to see what effect the above aspects can have on models similar to and beyond the one considered in this paper.

%%%%%%%%%%%%%%%%%%%%%%%%%%%%%%%%%%%%%%%%%%%%%%%%%%%%%%%%%%%%%%%%%%%%%%%%%%%%%%%%%%%%%%%%%%

\section*{Acknowledgements}
We thank Javier Rubio and Frank Wilczek for useful discussions. TT is funded by the Simons foundation. LV acknowledges support by the Vetenskapsr\r{a}det (Swedish Research Council) through contract No. 638-2013-8993 and the Oskar Klein Centre for Cosmoparticle Physics. TT and LV thank, respectively, the kind hospitality of Helsinki Institute of Physics (Finland) and Laboratori Nazionali di Frascati (Italy), where this work was initiated. The authors also thank the organisers of the Symposium in Honor of the Legacy of Vera Rubin, held in Georgetown University (USA), where this work was finished.

%%%%%%%%%%%%%%%%%%%%%%%%%%%%%%%%%%%%%%%%%%%%%%%%%%%%%%%%%%%%%%%%%%%%%%%%%%%%%%%%%%%%%%%%%%

\bibliography{axionHiggs}

%%%%%%%%%%%%%%%%%%%%%%%%%%%%%%%%%%%%%%%%%%%%%%%%%%%%%%%%%%%%%%%%%%%%%%%%%%%%%%%%%%%%%%%%%%

\end{document}